# High Quality Requirement Engineering and Applying Priority Based Tools for QoS Standardization in Web Service Architecture

C. Dinesh

***Abstract*— Even though there are more development to improving the QoS and requirement engineering in web services yet there is a big scarcity for its related standardization in day to day progress leading to vast needs in its area. Also in web service environment it always has been a big challenge to raise the standard of QoS in requirement engineering analysis. In order to meet higher level needs with high standard from the client's requirements, different services have been processed now a days depending on architecture evolvement through WSDL, SOAP and XML technologies. Even though it meets high level targets by implementing much more algorithm from previous suggested papers it is always a challenging task to reduce time consumption and appropriate response time from clients to the real world needs by providing high throughput as a result. So we have newly designed one different pattern oriented frame work for requirement engineering with web based services by applying our newly designed priority based tools for QoS standardization in Web Service architecture.**

.

***Index Terms*— QoS, Mapping, Priority Selection, Web Service, Filtration, User Request and RIA**

## I. Introduction

Quality of service is measured in many a manner for a software system based on its many requirements. Normally software systems are produced using more tools and are changed as application process for customer based on their requirements and are vastly used by an end user from its service provider. Here much more tools are used for this environment with the help of the interoperable machine-to-machine interaction over a network. In the web architectural sense WSDL-a machine readable language, UDDI (directory) and SOAP [1]-[8] are the primary tools for web based services.

C. Dinesh is a student of Mailam Engineering College at Villupuram under Anna University, Chennai, India .

Here SOAP is the protocol tool to communicate with machine to machine environment. Here we have to improve QoS in web service for requirement engineering. We design our architecture by following detailed cases for QoS web based environment. QoS, web services and requirement engineering are the important equipments for our proposed work in taking accurate results on verity of web based services for requirement services. These are improved using our proposed work. Thus our proposed work is designed using mapping of requirement to required services and secondly filtration is done dynamically for its quality of service.

## II. Improving Quality of Service

SOA is the basic form to produce one web based application at all. Here XML language is used as main tool for this service exchange between service provider and requester Service oriented architecture has special environment in its s/w as well as h/w specification for overall process in real time world for its day to day development. WSDL has the format of the interoperable language for any service based operation in web based area. Quality of service is here measured by introducing priority selection algorithm. It can be understood from our next definition and algorithm with detail. Quality of service [3] for one software system is improved under following environment according to user requirement. These are the conditions to be mainly considered for QoS through web based service.

### A. *Service Provider*

Service provider has the key aspect in all situations for service request. Based on the request received from service requester service provider has the concern and this process happens in SOAP based protocol which has platform independent operation and machine understandable script.

### B. *Service Requester*

Service requester leaves request to the service provider for his needs. All such needs are processed through web in a query format so that service providers can analysis and can response to the same.

### C. *WSDL Agreement*

This is web service definition language for explaining overall web services architecture for application communication. This is having the structure of XML based format with document or procedure oriented language. These have the capacity irrespective of different message format.

### *D. UDDI Registry*

Normally UDDI is used to communicate with SOAP protocol. All information about the web services are stored in this UDDI registry or directory.

### *E. Service Level Agreement*

In IT environment it is the process of agreement between one organization and stockholder or customer before continuing in advance for specified task or one formality or one white paper document in relevant area. It is a contract of service and also formally defined. Simply to say it is one type of agreement before starting the real services.

## III. REQUIREMENT ENGINEERING

Mainly changing the services happen in requirement engineering phase [6], and [8]. In order to satisfy individual or a group of people on the request it is reengineered on different environment related to the appropriate condition. While considering large software system process such as availability, changeability, testability, usability and security these all things have entire software architecture [9] for its thorough knowledge. There are different kinds of software quality attributes in the form of the above said.

### *A. Requirement Analysis*

The entire processes are checked for special requirements in relevant models. Here if there occurs any problem in that analyse then those problems are identified and resolved so that problems itself should not extend its presence, ie, in starting time itself, when it is found out is resolved with effect soon.

### *B. Requirement Management*

Managing the software systems and changes has important rules in requirement engineering. So before keeping in usage and process it is necessary to control each and every movement in system changes. If there is any such unexpected change then there must be an existing procedure for handling the same.

### *C. Requirement Validation*

Validation [10] checks for all system model requirements and documenting whether they meet sufficient modes from stakeholders and whether the same are consistent and dynamic in all cases with respect to any changes in its entire process.

## IV. MAPPING REQUIREMENT AS SERVICES

Here our main goal is to map the requirement phase to service through web services. All the requests are stored in a registry or database temporarily for mapping purpose [4]. When required request reaches the particular registry finally all those requests are mapped in appropriate position for further division like filtering tool process. This mapping is the first work to be done before filtering all received request from the user. These can be explained through mapping algorithm.

<u>Data entering and storage algorithm</u>

```
Procedure
    //entering the request
    //receive the request
    GetMsg = SelectSource+MsgReceived
    StoreMsg = MsgReceived+NewMsg
    SelectSource < - n(n_{i+1},n_{i+2}……..n_{i-n})
    MsgRecived = InputOfMsg(i)
    NewMsg = EnteredMsg(i)
    if (GetMsg == Nil)
    //proceed to database storage area for searching
    StoreMsg = NewMsg(i)
    //upto the level of user request
    else GetMsg = SourceMsg(i-ExistingMsg)
    //Continue upto the existing level
    MsgReceived = SourceMsg(i)+NewMsg(i)
End Procedure
```

<u>Mapping Algorithm for Service Requester:</u>

```
procedure
    //Select "z' group up to (0,…n-1) level
    //choose required sets for request
    //let i be the initial message
    //from service request
    if(z == i)
    z => (0,….,n-1)
    for each i, j<-1,n do
    for first selection i<-1, n-1
    get request "i+1"
    increment the value
    n = assignment of request+i
    else
    {
    z = i_{i+1,}i_{i+2}………i_{i+n}
    //for entire request
    }
    end for
    end for
    //store all the values and
end procedure
```

Mapping the request [8] for its next process is the first step in software system process and the next process starts with filtering from successive work continuously. Then proposed work, namely priority selection algorithm is explained in detailed manner with algorithm. Service requester depends on service provider for response from service provider. This response is provided based on high priority method from requested string from user level. When related string is received from data base normally all types of related strings will be outcome based on the request. It is a normal event happening in response time. Apart from its existing tendency we have designed different priority selection algorithm. In our paper priority selection algorithm is a state of art and no new concept is there like our aim from what we tell in this paper.

## V. Filtering of User Request and Algorithm

Filtering is the process as soon as the requests are mapped from user to the data base as minimum reduced data sets so that the quality of service can be improved in a high level with high throughput. Firstly the necessary strings are selected based on the input values so that the level of unwanted request can be reduced from user. Then these datum are ordered in a sequence way for further movement. QoS Architecture with Web Service Elements

Service oriented architecture [2] explains its way of nature from below diagram. SOA has combined architecture for its QoS and other mechanism. User sends the request to the service provider for his needs. Those requests are stored as one frame formation in one data base directory for reprocessing. These frames are then considered for further division from that particular position.

When there is gathered a great quantity of requests are stored automatically. Then they are reprocessed from original frame form for next level, ie the mapping procedure. These processes are next considered to the nearby filtering tools. This filtering tool takes care of ordered set of request.

As final process the proposed “priority selection algorithm” does its working perfectly to the required level. It works on the basis of higher priority mode from the filtering point. This work is done effectively to separate needed portion of string from the request data base. This process is explained in clear manner by specified algorithm. Service oriented architecture has the following methods for doing the job from service requester to service provider and vice versa. The main components are classified as in following components

- User
- Request database
- Mapping of the request
- Filtering of mapped message
- WSDL and UDDI
- SOAP
- Cloud architecture or SOA

The above classified things have more explanation in this paper for further details.

Filtering Algorithm for Request

```
procedure
    //Let “S” be Storage area for filtered element
    //”T” the total number of messages & i=1
    //Input n and i for element and counter
    //S={ni,ni+1,……..,(ni-n-1)}
        if(n<=0)&&(n == i)
            for(i=0;i< = n;i++)
              for(i=0;i< = n;i++)
                  S=i
                  S++
              End for
            End for
        End if
    // otherwise n++;
stop the program
```

WSDL Script Model

```
<wsdl:binding name="nmtoken" type="qname">*
<wsdl:documentation .... />?
<-- extensibility element --> *
<wsdl:operation name="nmtoken">*
<wsdl:documentation .... /> ?
<-- extensibility element --> *
<wsdl:input> ?
<wsdl:documentation .... /> ?
<-- extensibility element -->
</wsdl:input>
<wsdl:output> ?
<wsdl:documentation .... /> ?
<-- extensibility element --> *
</wsdl:output>
<wsdl:fault name="nmtoken"> *
<wsdl:documentation .... /> ?
<-- extensibility element --> *
</wsdl:fault>
</wsdl:operation>
</wsdl:binding>
```

SOAP script

```
<?xml version="1.0"?>
<soap:Envelope
xmlns:soap="http://www.w3.org/2003/05/soap
envelope">
<soap:Header>
 </soap:Header>
 <soap:Body>
<q:MessageNeeded
xmlns:q="http://www.example.org/message">
<q:MessageID>TTTTT</q:MessageID>
  </q:MessageNeeded>
</soap:Body>
</soap:Envelope>
```

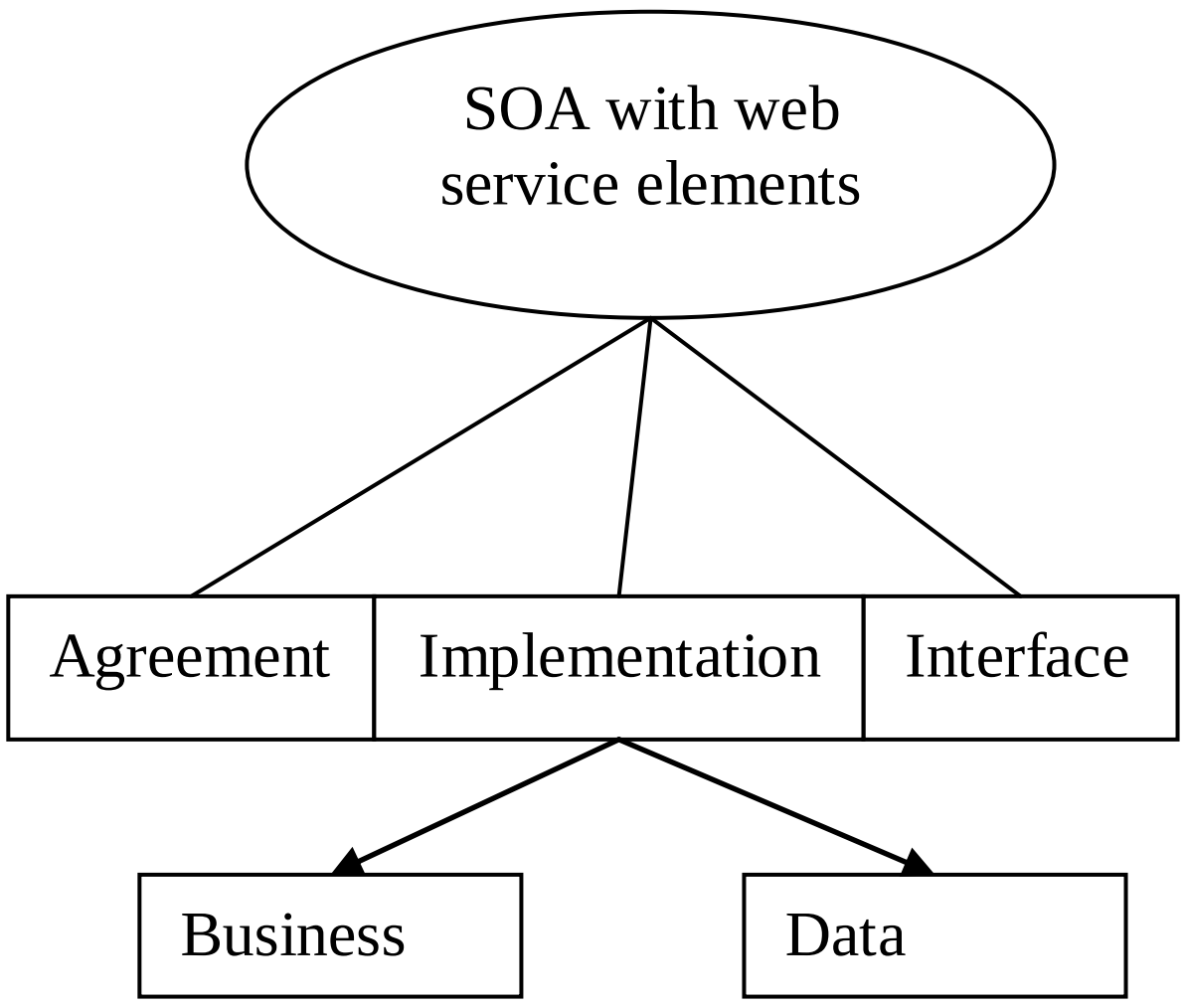

Fig.1 Component of SOA

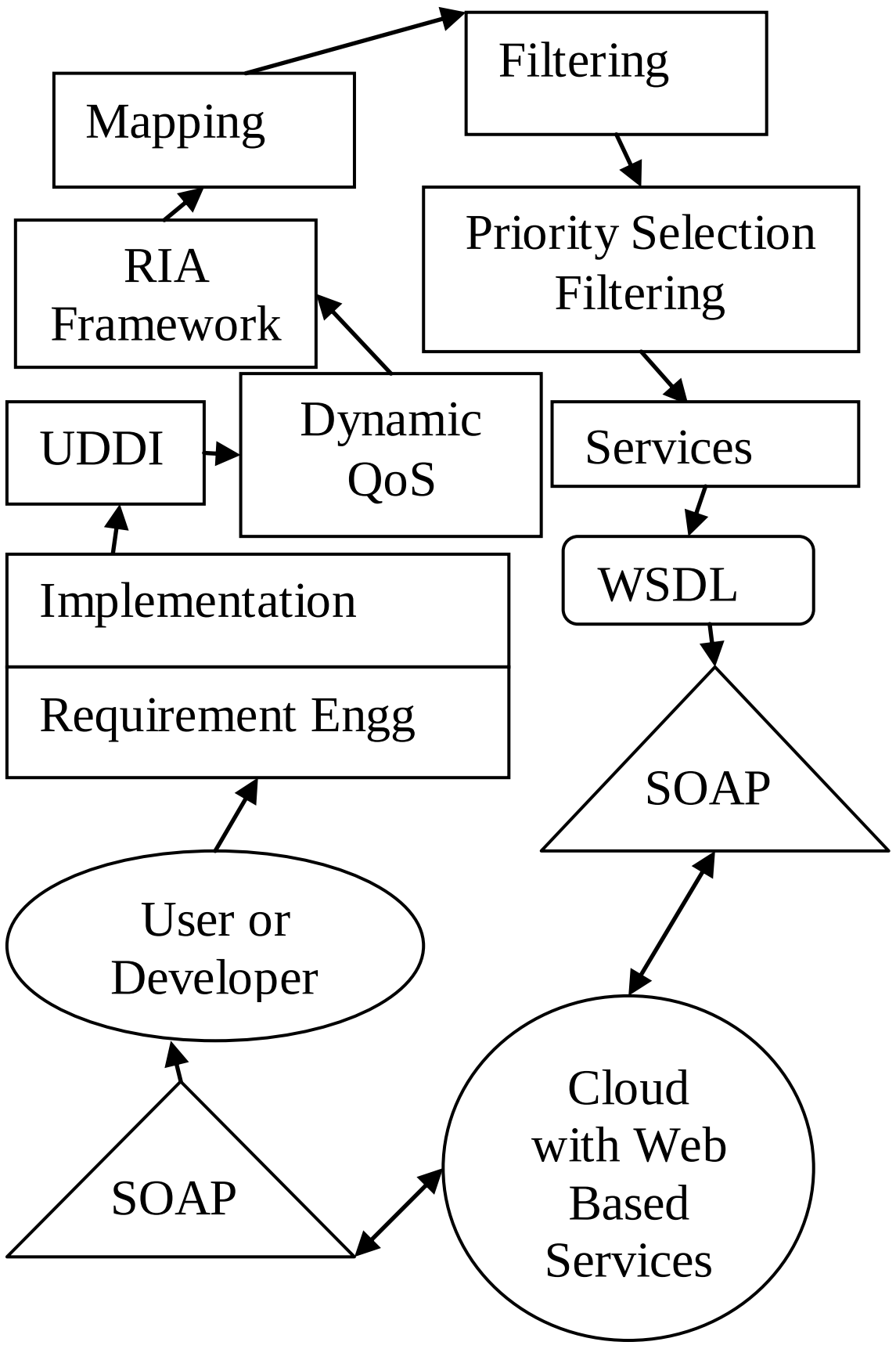

Fig. 2 SOA using web services for requirement engineering

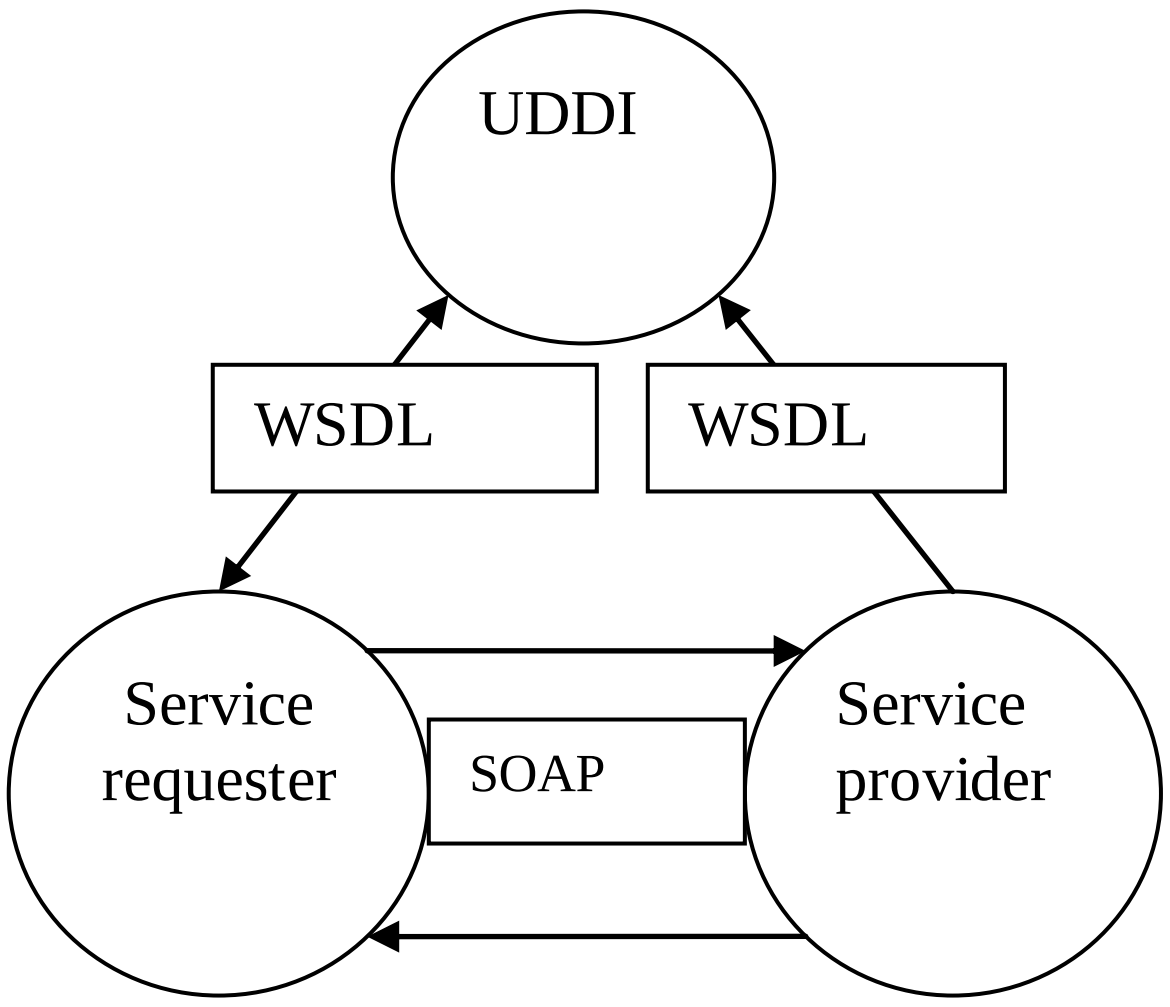

Fig. 3 Web service architecture

## VI. Requirement Identification Analysis

Here we concentrate on step by step activity in web service environment for RIA. We implement our proposed scheme by RIA frame work. So we design an algorithm for filtration providing quality of service particularly in dynamic environment. Priority selection algorithm is an efficient one to filter the level of required string from the request side, ie database for all requests. These works are separated from mapping, filtering and finally to the priority selection level in a step by step process respectively. By reprocessing all these steps the required processes are achieved well in accurate level. By accessing reengineering models all system components are combined in its final attempt by utilizing the WSDL and UDDI registry.

The SOAP messages [3], and [7] act from both levels, ie, service request and service provider for machine to machine interaction. SOA has three types of stages such as enterprise service layer, domain service layer, and application service layer. Normally SOA is a set of procedures designing and developing the software in the form of interoperable services. The following mathematical aggregation formula explains the overall priority selection for request from database. Required string= (Request from database-Mapping overall request-Filtering of overall request)/3

- D= Database request
- M= Mapping of request
- F= Filtering of request
- In mathematical terms, it has as ,

$((D_{i+1},D_{i+2},D_{i+3}\ldots\ldots D_{i+n})-(M_{i+1},M_{i+2},M_{i+3}\ldots\ldots M_{i+n})-(F_{i+1},F_{i+2},F_{i+3}\ldots\ldots\ldots F_{i+n}))$/total aggregate of request. Thus overall priority selection is reserved in final process. Here for clarity of mathematical calculation, we can select the following methods for priority selection. S= (D-M-F)/3 is the reduced or brief form of the formula defined.

## VII. Advantages of Priority Selection Tools

Even though there are many implementation concepts available regarding with the service selection algorithm in service oriented architecture yet there is no such algorithm related to priority selection based one as we have provided in this paper. As far as our proposed algorithm is concerned it is all about to minimize the time efficiency and entire throughput of immediate available of services. Mainly we focus on the immediate response for selected quarry based information while considering one application loaded into the SOA level or cloud architecture. When required request is received from the mapping immediately It is processed to the next level and then it is followed by filtering algorithm and finally our proposed work “priority selection algorithm” does its working for its final process to filter requested information from service provider again to service requester as output based on one’s needs. Normally proposed work is preceded in as many ways for its thorough string filtration.

While comparing our proposed work with any other related work mentioned in this paper, others have not used any such algorithm that we have presented here for step by step request filtration as we have handled here by our algorithm.

Proposed Priority Selection Algorithm

```
procedure
    //Consider for an application
    //Input the element
    //Input Selection Tool Variables
    //Give high priority to the element to be
      searched
    //Sort the element according to algorithm
    //Consider low priority element to reserve
    For each filtering elements
            I < - (high priority element(n) < -
                 filtering            element)
            I < - i_{i+1},i_{i+2},............i_{i+n}
    //separate all filtering elements
    //Input p, q, r, & s respectively for sorting
    element such as high, low, very low, medium
    if(i< = 0)
    {
    //p= higher priority of element
    //q=any string to be searched
    //r, s=sample message strings for input
    p - > high priority element
    }
    q - > high priority element
    else if (p == q)
    {
    p = high priority (q) < - searched element(r)
    s = p_{i+1},p_{i+2}.......p_{i+n}
    }
End Procedure
```

XML Quries

```
<?xml version="1.0" encoding="ISO-     8859-1"?>
<service requestid="2345345"
xmlns:xsi=http://www.w3.org/2001/XMLSchema-
instance
xsi:noNamespaceSchemaLocation="servicerequest.xsd"
>
 <requester>vvvvvvvv</requester>
 <serviceto>
  <portname>tttttttt </portname>
  <ipaddress> rrrrrrrrr</ipaddress>
  <recport>45665677</recport>
  <country>uuuuuuuu</country>
 </serviceto>
 </service>
```

## VIII. REQUIREMENT ANALYSIS

Normally developer or user has to provide service on the basis of requirements from various needs from different clients or request. Any delivery of services to the clients must meet all system requirements [1] under usable and stable condition. So according to this analysis we have designed our proposed system. This process is based on a few categories of requirement engineering such as

- Quality model
- International standards
- Quality factors and sub-factors
- Completeness

Now we can see some detailed explanation about these things in requirement engineering.

### A. *Quality Model*

Quality model is explored in business context as software functional quality and structural quality. Normally software functional quality tells about functional elements of systems or softwares. Mainly functional model comes from inputs and behaviours as well as outputs. In short it defines use cases of the systems. Here the functionality of the system is described from use cases. For example these functions can be technical element, calculation and a few system interaction oriented process. In non functional categories, it has vice versa operation to functional element as this has operational function of the system. The non-functional elements come under the quality of the system. This quality of systems includes attributes and goals.

### B. *International Standards*

All standard software quality is checked whether it belongs to individual development or endeavour purpose or belongs to any defined categories such military or any governmental purpose, etc. It is a standard for defining the step by step process such as requirement, design, implementation, verification, and maintenance. These things are explained by software life cycle.

### C. *Quality Factors and Sub-factors*

It has the non-functional cases and includes understandability, conciseness, portability and document. The first thing tells that the software product must be clear in all level. All design and documentation pattern must be clearly written. The second thing concentrates the line of codes.

### D. *Completeness*

It is one type of process to check if the particular software product is completed in it all level such as line of code, input methodology, pattern development, and sub routine elements.

Now we need to analyse three levels of user requirements for priority selection such as

- Normal requirements
- Expected level requirements
- Exited level requirements

### E. *Normal Requirements*

According to normal requirements when client searches for the required elements he gets results in minimum level basis. As no maximum filtration is applied to this level when he searches for appropriate elements he will not have strong output for that. Because the filtration algorithm does not make any sensible operation for client's search criteria he can't get appropriate elements for his search results.

### *F. Exited Requirements*

This level of requirement does a little bit of stronger result than the previous one. When the needed jobs are done for best result the related results are received in its entire position. The major drawback of this requirement is time consuming.

### *G. Expected Requirements*

This level of process is better option using our proposed algorithm. Since it explores our expected level of outputs time consumption and system throughput are entirely improved so that client can be successful for his expected level of result. We can understand these results from our proposed algorithm "priority selection" and its process.

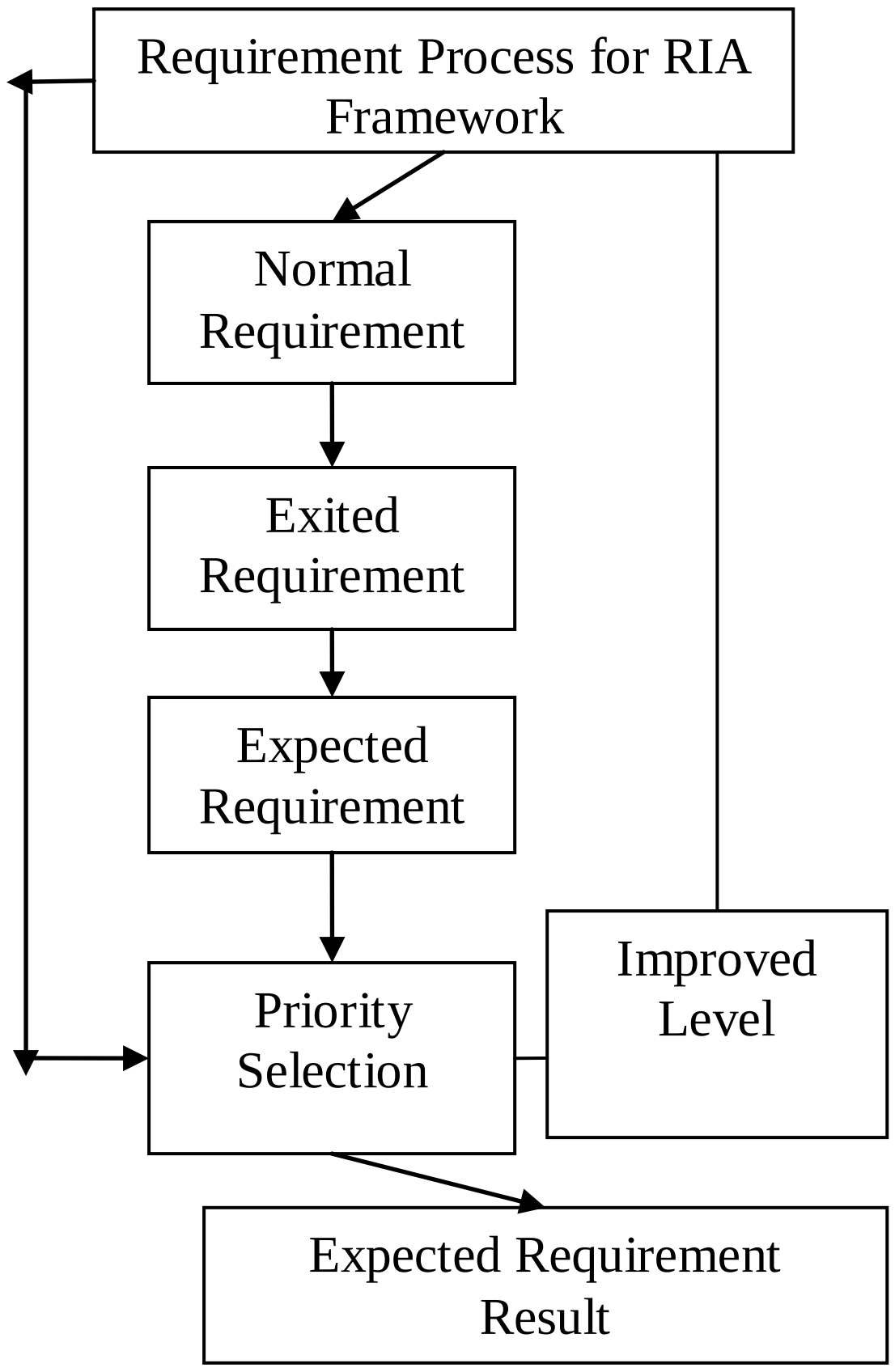

Fig. 4 RIA framework process with accurate result

## IX. Related Works

Our paper clearly tells proposed concept in a detailed manner with performance evaluation. We have defined one tabular column in performance analyzing for our step by step process in SOA manipulation for request filtering. Paper [1] tells only about service registry in web service selection. Here nonfunctional properties are explained in brief manner. Validations are maintained in each level. Paper [2] explores functional and use case requirements to its credits. It mostly tells the interior design of the system behavior. All use case scenarios are clearly mentioned. Paper [3] tells service based service selection approach for service selection. Mainly this paper studies functional based, nonfunctional based and user based service selection process. Paper [4] tells about requirement engineering and its development with mapping concept. Paper [5] tells non-functional requirements and its impacts in software engineering. Paper [6] explores non-functional elements for architectural decisions and validation of system architecture. Paper [7] explores ideas about response time, efficiency, round trip time of QoS using web services distribution management. It tells how to give good service to web by load balancing and finally paper [8] analyses QoS in web services and deployment of application in cloud manager. Thus each and every paper explores ideas about above mentioned things related to our nature of things which elaborately have said through our algorithm.

## X. Performance Analysis for RIA Framework

Thus our study on priority selection algorithm analyses how to provide the top level value to the request for easiest merging for required resources so that particular service only can reach to the destination route or party on one's request based on accurate queries. The following table compares some requirements based on RIA framework.

TABLE 1
PERFORMANCE ANALYZING FOR RIA FRAMEWORK

| **S.NO** | **Proposed Algorithm Improvement for RIA framework** | | |
|---|---|---|---|
| | **Mapping** | **Filtering** | **Priority Selection** |
| 1 | receiving request from the clients | initialize the element | algorithm based filtering |
| 2 | vast kind of unfiltered element | sorting for the filtering | selection based tools |
| 3 | related database as request from end user | database checking using tools | receiving the filtered element |
| 4 | searching of related and relation with filtering tools | ordering and sequencing | searching only for related data |
| 5 | raw data receiving and gathering | ready to send | selection of high priority element |

Our real time work is done by collecting sample from a large web based searching for customer related things from many web sources. The following sample collection output explores the best results from our proposed methodology for related searches so that the best services can be given based on priority algorithm. By our algorithm results we can understand the efficiency of filtering results and final graph improvement. In order to get good improvement for our proposed algorithm we have handled some tools for analyzing our proposed scheme with efficient manner so that the required customer satisfaction can be achieved.

## XI. Sequence of Proposed Work Process for RIA

Here we design our entire work as many category for its proper order for quality of service. The following are the phases in our RIA frame work architecture process.

- RIA- requirement identification framework
- Mapping - changing requirements as services
- Filtering - for QoS (dynamically)
- Priority selection- proposed work

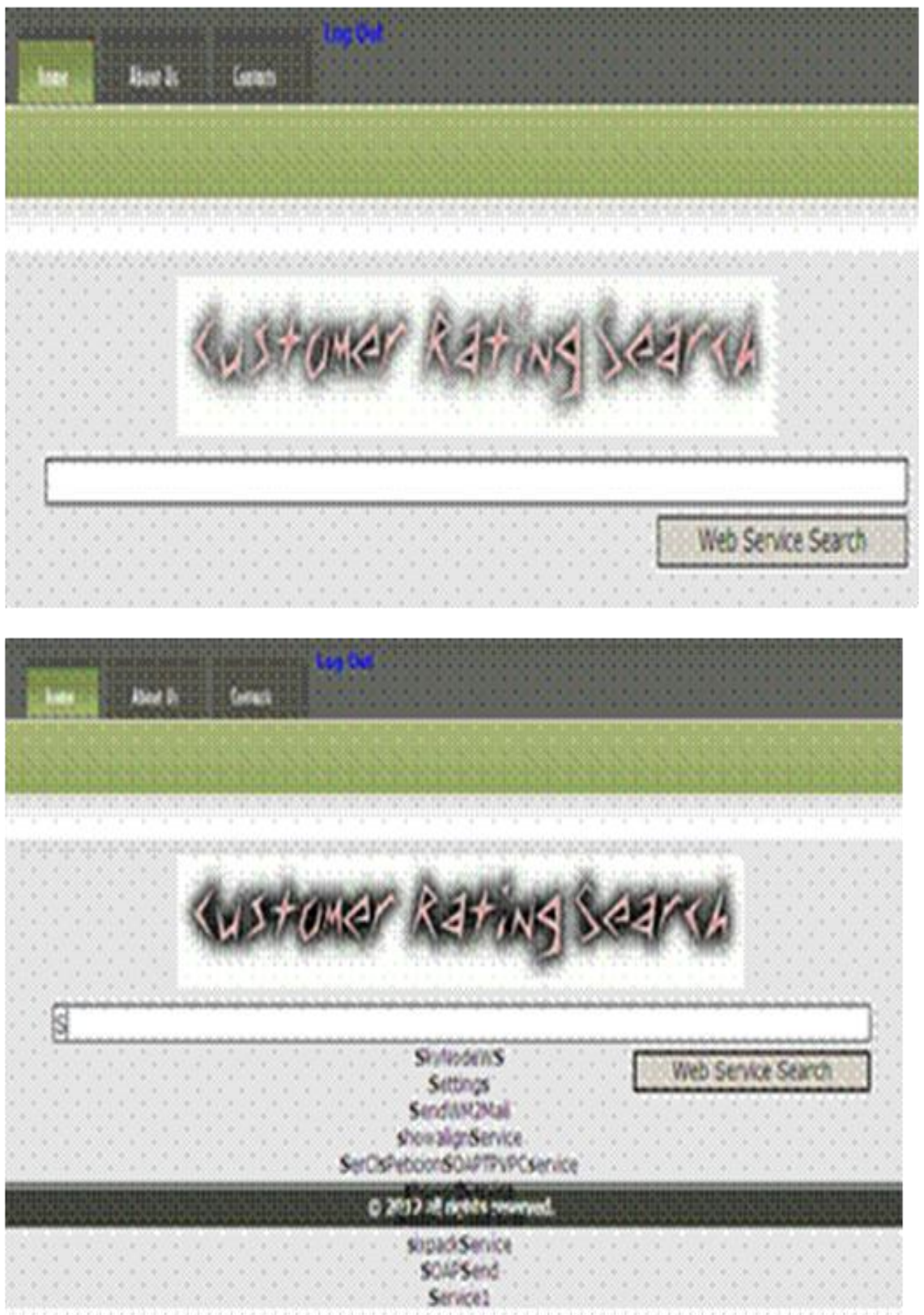

Fig. 5 Searching for related strings

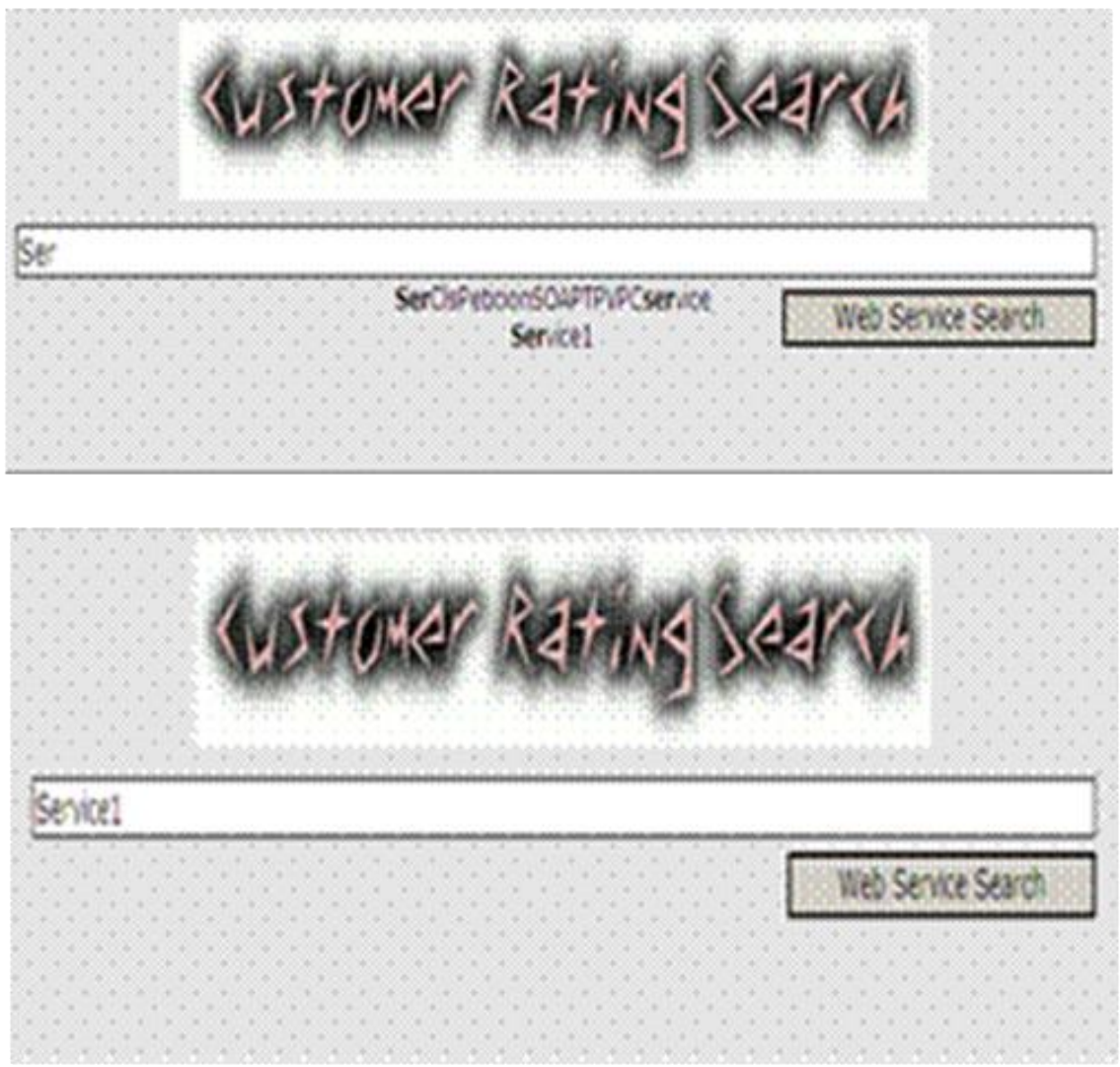

Fig. 6 Entering and searching for related services

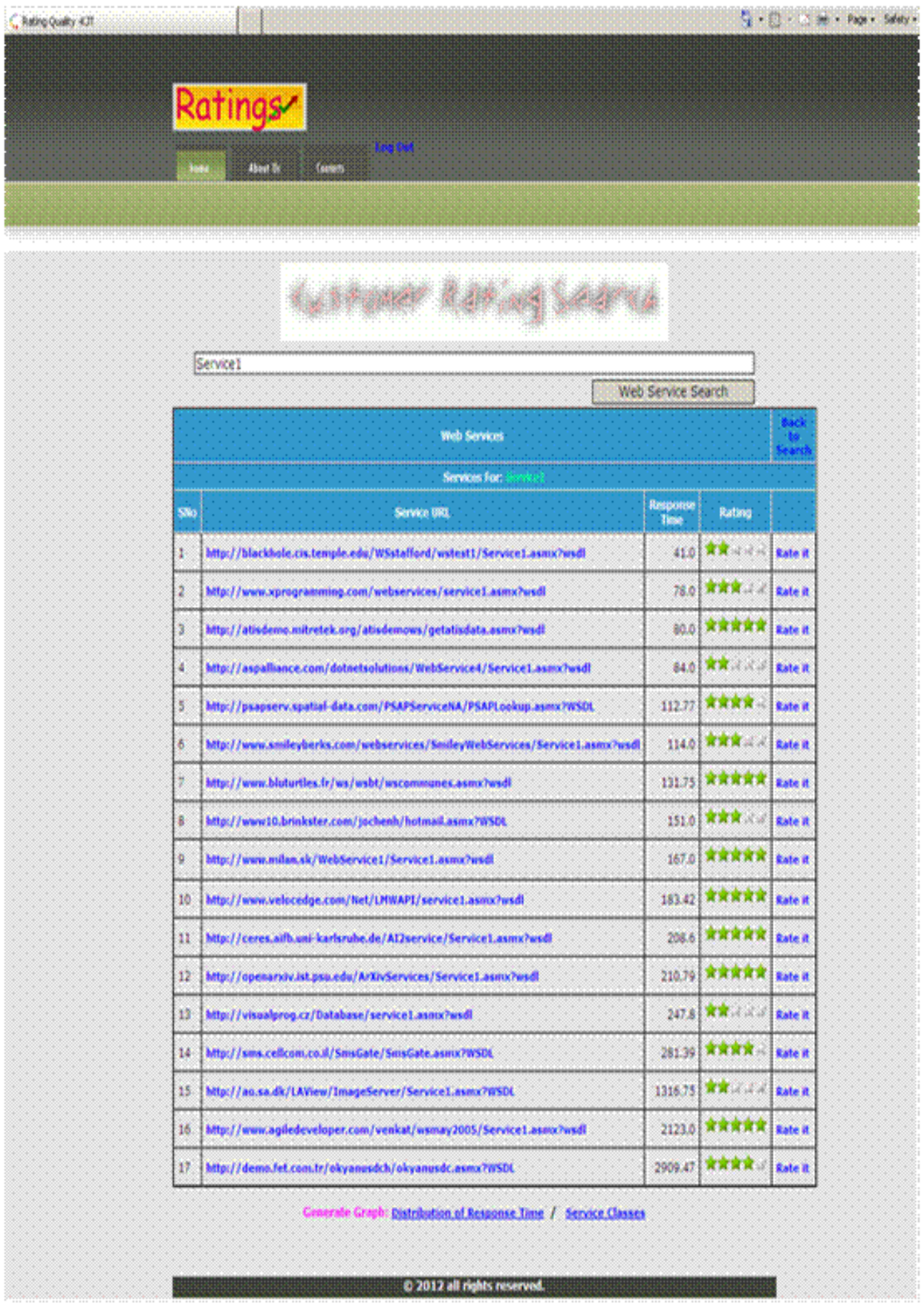

Fig. 7 Web service samples for request

## XII. Dynamic QoS with Web Services

Initially we start work from requirement engineering level where changes have been done before the requirement engineering phase itself. Now as second part we do our implementation for required services. All our requests are processed from UDDI registry and then Quality of services is followed by dynamic environment for our convenient. From here our three level of filtration starts its working. Now requirements are mapped to required services. These are our step by step changes and process for requirement engineering using web based services for user level request. Figure 5 and Figure 6 show that when the required strings are given as input to the web services in requirement process or engineering in its initial stage the required strings are automatically stored in its database. Figure 7 explores required result for a vast kind of web site based service. Now performances are measured in a graphical format for any such related types as we have here in our result. Our graph has high level customer rating range for customer requirement. No paper has our final output result in this way by graphical measurement. For our requirement engineering process we have chosen a few real time web services from many web sources. Also we have used the java language for our output analyzing.

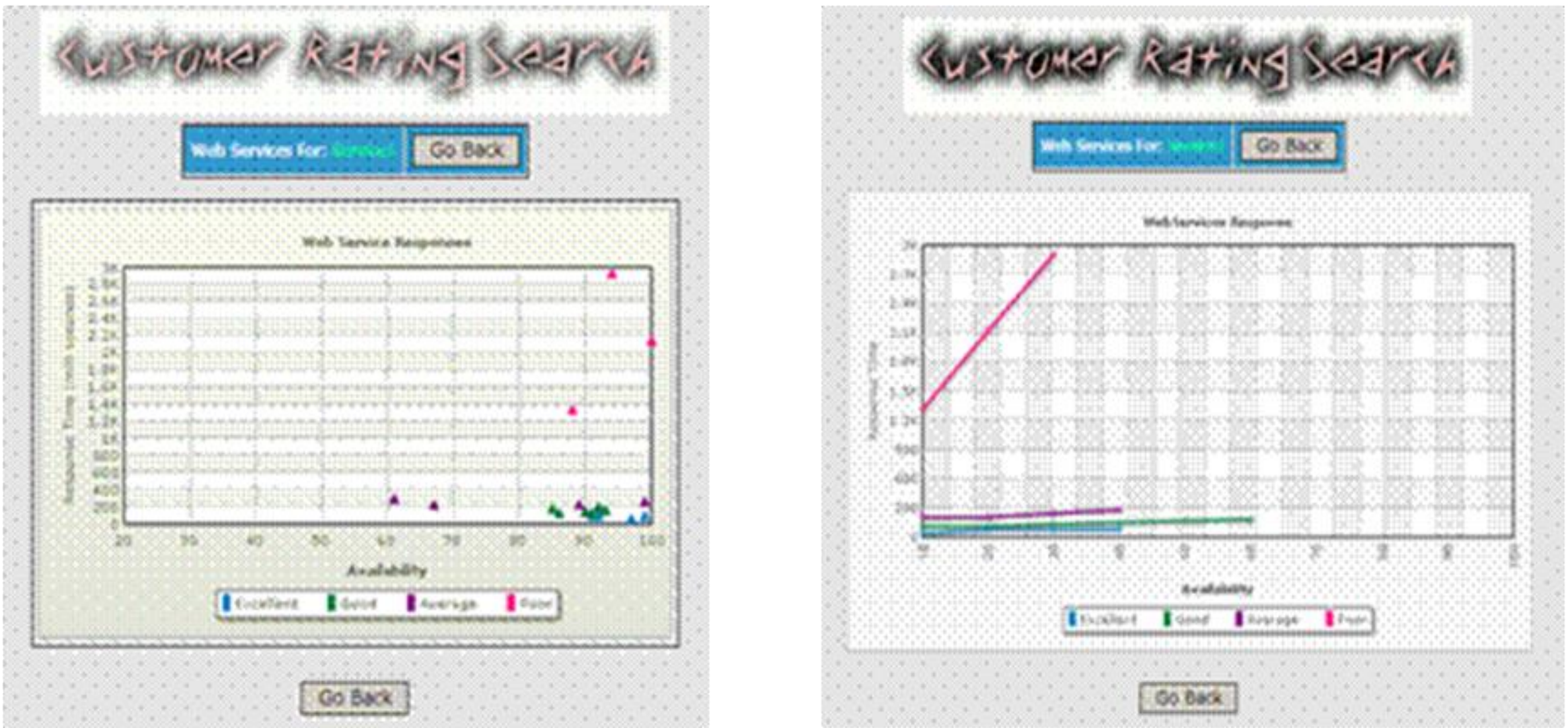

Fig. 8 QoS improvement graph by our analysis

Thus we have entirely improved our final proposed architecture so that the best results can be achieved compared to related papers in requirement engineering processes or identification. Our graph identifies its improvement in a high level from customer rating. Thus customer rating gives best output result according to our final implementation. Normally there is happening much more research process for such customer rating using software applications [10] in useful manner for elaborated results now a days . Normally customer requirements must be testable, measurable and also must meet business requirements for its good standards. But in our paper we have not covered business purpose since it is not our scope. Even though there is more requirement analysis processes under research security and privacy [11] are also major concern and threats to the software engineering domain forever as far as its private environments and privacies are concerned. But it is apart from our entire scope.

## XIII. QoS Evaluation after PSA

Service oriented architecture has the structure of figure 1 from our drawn one for its entire process based on our proposed algorithm. When our proposed method works its filtration process, immediately it takes request from filtering database and searches for required data string in correct location by its sorting process. Thus it uses higher priority tool for sorting work. As the searching tool does its working from appropriate string identification it can get its success criteria without any break and stoppage within the specified time. So by our proposed algorithm the request to be searched has high receiving result. From this type of process system process and its user request can be improved to a high level so that the throughput of system waiting time is maintained well and improved to such a great extent.

## XIV. Real Time Sample Result

Thus by taking some real time based website models we have handled our entire design for improvement of request filtration as soon as one tries for particular requirement from given input. Our algorithm tells clearly about how to filter unwanted and unrelated strings from data storage for further process and how to avoid needless strings to save user’s time and how to regularize all input methods for accurate results by each filtration method and how to provide high priority to the required input within requested portion of string. So all the above fulfillments are satisfied by new formula that we have derived and moreover our results based on real time examples give proper result for our proposed aim. This can be seen by all figures that we have given here and taken as sample output from implementation in this paper. Here we have elaborated our full details in vast manner in this paper with step by step algorithm including some sample real time output including the throughput of graph. Even though there is a different approach like our referenced model in this respect for some related expectation our paper result based on priority selection based algorithm provides efficient outputs from real time environment to outside world.

## XV. Evaluation of Graph Rating

Our graph explores some idea about customer based ratings for particular web based result in all comparison with two graph models of above for expected result in requirement engineering process. Thus all our experiment result tells sharply by taking examples from a few website models and its customer ratings for request by user queries in efficient manner using priority selection concept. Hereafter figure 8 tells about quality of service improvement in requirement engineering and its requirement analysis identification. From left and right side graph we can understand the tendency of service availability in web services for particular option by user. Services are rated using the scalability level of excellence, good, average and poor by the web service management in requirement analysis identification for availability of customer services. Based on the request of the user in the web service repository everything has to be measured to know about its excellence of filtered input for the best customer ratings. Thus our graph result tells all such improvements by each step by step filtration process that we have done already from user input for web content. The level of graph gives some excellence, average, and poor performance in each filtration process for required level of queries. Thus ease of availability of services is measured using graph for its through analyzing. The right side graph tells clearly about good performance of availability of services in a correct path with good throughput based on input from web services to user. On the other hand left side graph does not show its tendency so sharply compared to right side one in service availability. This can be seen from its straight line path with different level than that of the left side graph in which the same availability of services has different format.

## XVI. Conclusion and Future Work

Thus we have implemented RIA (Requirement identification analysis) frame work in validation for requirement engineering. Our paper analyses the entire SOA for web based service request process and frames clearly with required algorithm for the selection of our proposed "priority selection algorithm" methods for best validation . Our proposed scheme works well for accurate result in providing specified service in SOA based environment. Here we have analyzed overall performance of web service result using our proposed scheme by best outcome since we have received immediate response. Thus we have improved the quality of web services in our efficient proposed algorithm. From our proposed algorithm and result we can understand the overall web based quality of service and we have here clearly measured QoS using our proposed scheme. In initial stage changing services happen in requirement engineering phase itself. Above all QoS has dynamic level environment in our design. Thus we prove requirement engineering phase in a high level for quality of service in web based services in efficient manner using priority selection algorithm with required implementation process. As a future work we have planned to work these environments for requirement engineering using advance web based search process for immediate result to user request in big cloud based concept to meet current requirement engineering process. Thus we have measured the final results by taking real time sample of a few websites, ie, we can see figure 7 "web service samples for request". This figure explores idea about the suitable web based result by searching for best filtration output. For this However we have designed one efficient RIA framework with high throughput for requirement engineering phase through web based service with proper result analysis from our application in our paper.

## References

[1] Ruth Malan and Dana Bredemeyer, Bredemeyer Consulting, Functional Requirements and Use Cases, http://www.bredemeyer.com, march-2001.

[2] M. Sathya, M. Swarnamugi, P. Dhavachelvan & G. Sureshkumar, Evaluation of QoS Based Web- Service Selection Techniques for Service Composition, (IJSE), Volume (1),- 2011.

[3] Kyriakos E. Kritikos, QoS-based Web Service Description and Discovery, December- 2008.

[4] Dongyun Liu, Hong Mei, Mapping requirements to software architecture by feature-orientation, Requirement Engineering- 2003.

[5] Lawrence Chung and Julio Cesar Sampaio do Prado Leite, On Non-Functional Requirements in Software Engineering, Springer- 2009.

[6] Ruth Malan and Dana Bredemeyer, Bredemeyer Consulting, Definigh Non-Functional Requirements, http://www.bredemeyer.com, march-2001.

[7] G. Vadivelou, E. Ilavarasan, R. Manoharan, P. Praveen, A QoS Based Web Service Selection Through Delegation, International Journal of Scientific & Engineering Research Volume 2, Issue 5, May- 2011.

[8] David Villegas and S. Masoud Sadjadi, Mapping Non-Functional Requirements to Cloud Applications, IPDPS Workshops- 2010.

[9] http://www.sei.cmu.edu/publications/documents/95.reports/95.tr.021.html, Barbacci, Mario; Klein, Mark; Longstaff, Thomas; & Weinstock, Charles. Quality Attributes (CMU/SEI-95-TR-021 ). Software Engineering Institute, Carnegie Mellon University, December- 1995.

[10] http://www.ifi.uzh.ch/rerg/people/koziolek/publications.html, Software Architecture Optimization Methods: A Systematic Literature Review, Aldeida Aleti, Barbora Buhnova, Lars Grunske, Member, IEEE, Anne Koziolek, Member, IEEE, and Indika Meedeniy, September-2012.

[11] Bijwe, Ashwini; & Mead, Nancy. Adapting the SQUARE Process for Privacy Requirements Engineering (CMU/SEI-2010-TN-022). Software Engineering Institute, Carnegie Mellon University, July- 2010.